\documentstyle[aps,multicol,epsf,epsfig,axodraw]{revtex}
\def\axoscale{2.0 }
\def\axowidth{0.3 }
\def\ruleright{\vspace{-1.5\baselineskip}\begin{multicols}{2}\ \linebreak\raisebox
{-2.45mm}{$\lceil\!$}\hrulefill\end{multicols}\vspace{-\baselineskip}}
\begin{document}

\title{Complex Formation Between Polyelectrolytes and Ionic Surfactants}
\author{Paulo S. Kuhn, Yan Levin\footnote{Corresponding author; e-mail: levin@if.ufrgs.br} \, , and Marcia C. Barbosa}
\address{
Instituto de F\'{\i}sica, Universidade Federal
do Rio Grande do Sul\\ Caixa Postal 15051, CEP 91501-970
, Porto Alegre, RS, Brazil}
\maketitle
\begin{abstract}
The interaction between polyelectrolyte and ionic surfactant 
is of great importance in different areas of chemistry and biology. In this 
paper we present a theory of polyelectrolyte ionic-surfactant solutions. The 
new theory successfully explains the cooperative transition observed 
experimentally, in which the condensed counterions are replaced by 
ionic-surfactants. The transition is found to occur at surfactant 
densities much lower than those for a similar transition in non-ionic 
polymer-surfactant solutions. Possible application of DNA surfactant 
complex formation to polynucleotide delivery systems is also mentioned.

\end{abstract}

\bigskip

\begin{multicols}{2}
\section*{\bf Introduction}

Solutions containing polyelectrolytes remain an outstanding
challenge to physical-chemistry. Due to the long-ranged nature 
of the Coulomb force our understanding of this class of polymers 
is still quite rudimentary. This situation can be compared to the 
one that existed in electrochemistry of the turn of the century, before
Debye and H\"uckel (DH) presented their, now famous, theory of strong 
electrolytes \cite{DH}. The fundamental question that must be addressed 
by any successful theory of polyelectrolytes concerns with the role 
played by the counterions. In this respect, the traditional theories 
of liquid state are not of great help, since most of the approaches
based on resolution of integral equations come to a dead-end when 
the numerical schemes used to tackle these difficult problems fail
to converge. The scaling theories, which have been so successful 
in elucidating the properties of nonionic polymers \cite{DeGennes}
have, so far, proven futile in the face of large number of length scales
relevant for polyelectrolyte solutions. What seems to be lacking 
is a mean-field theory of polyelectrolytes similar to the one 
created by Debye and H\"uckel for simple electrolytes, and Flory
for non-ionic polymers. In our previous work we have attempted to
construct such a mean-field theory for one special class of polymer 
solutions, the rigid polyelectrolytes \cite{LevBarb,Rod}. The constraint 
of rigidity allowed us to study the effects of electrostatic 
interactions decoupled from that of conformational structure 
of polyions. The theory has proven to be successful in elucidating 
various thermodynamic properties of rigid polyelectrolytes in the presence, or 
in the absence, of monovalent salt. In this letter we shall present a theory of 
rigid polyelectrolyte and ionic-surfactant solutions.

The interaction between polymers and surfactants is of great practical  
importance in areas as diverse as colloidal stabilization, polymer 
solubilization, mineral flotation and flocculation, as well as various aspects 
of molecular biology and biochemistry \cite{Haya}. In many practical 
applications the polymers are dissolved in some sort of polar solvent, typically 
water, leading to monomer ionization. This situation is very common for 
biological systems. For example, in an aqueous solution, the phosphate groups 
of a DNA molecule become ionized, giving it a net negative charge. Similarly 
the phospholipids, which compose the cell membrane, in aqueous environment 
acquire a net negative charge. The repulsion between the like-charged 
molecules makes the introduction of a polynucleotide sequence into a 
cell a formidable challenge to molecular biologists. It has been observed, however, 
that in the case of binding by ionic surfactant dissolved in a polyelectrolyte 
solution, the adsorption isotherms show a striking degree of cooperativity. This 
surprising phenomenon suggests that ionic surfactants or ionic lipids can be 
used as a ``packaging" in order to deliver polynucleotides into living 
cells. Indeed, some recent experiments demonstrate that the cationic lipid 
reagents provide some of the best methods available for the gene delivery 
systems \cite{Felg}.

\section*{\bf The Model}

The solution  under consideration consists of anionic 
polyions, monovalent salt, and cationic surfactant, inside a volume $V$ 
(see Fig.$1$). It is  important to remember that the overall 
system is charge neutral, what implies that the negative charge 
of polyions and the positive charge of surfactants is 
counterbalanced by an appropriate number of counterions 
(univalent cations), and coions (univalent anions), respectively. Furthermore, to 
simplify the analysis we shall assume that all of the counterions are 
identical, whether they are derived from polyions or from disassociation 
of monovalent salt. A similar approximation will be made in the case of coions.
  
In order to study the interaction between an ionic surfactant and a 
polyelectrolyte we resort to the simplest possible model. The rigid 
polyions, of density $\rho_p$, are represented by cylinders of 
length $L$ and diameter $a_p$. Each polyion has a charge $-Zq$, uniformly 
distributed along the length of the cylinder. The spacing between each 
charged group is $b \equiv L/Z$. The cationic surfactants, of density 
$\rho_s$, are modeled as flexible chains of $n_s$ monomers each, with the 
head group carrying a charge $+q$. For simplicity we shall assume that 
each monomer is a sphere of diameter $a_c$. The density of counterions 
(cations) is $\rho_{count}=Z \rho_p + \rho_{salt}$, while the density of 
coions (anions) is $\rho_{coion}=\rho_{salt}+\rho_s$. Both the coions and 
the counterions will be modeled as hard spheres of diameter $a_c$ and charge 
$\pm q$ located at their centers. The solvent (water) will be represented by 
a uniform medium of dielectric constant $D$.

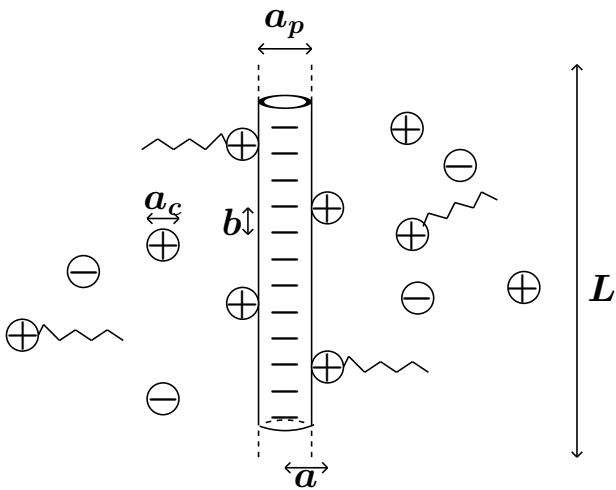
\begin{figure}[h]

\vspace*{2cm}
\begin{center}
\begin{picture}(70,70)(0,0)
\Line(20,10)(20,70)
\Line(30,10)(30,70)
\DashLine(20,3)(20,10)1
\DashLine(20,70)(20,77)1
\DashLine(30,3)(30,10)1
\DashLine(30,70)(30,77)1
\Oval(25,70)(1.2,4.5)(0)
\DashCArc(25,-4)(14,75,105)1
\CArc(25,22)(14,250,293)
\Text(25,10.5)[c]{\Large\boldmath$-$}
\Text(25,15.5)[c]{\Large\boldmath$-$}
\Text(25,20.5)[c]{\Large\boldmath$-$}
\Text(25,25.5)[c]{\Large\boldmath$-$}
\Text(25,30.5)[c]{\Large\boldmath$-$}
\Text(25,35.5)[c]{\Large\boldmath$-$}
\Text(25,40.5)[c]{\Large\boldmath$-$}
\Text(25,45.5)[c]{\Large\boldmath$-$}
\Text(25,50.5)[c]{\Large\boldmath$-$}
\Text(25,55.5)[c]{\Large\boldmath$-$}
\Text(25,60.5)[c]{\Large\boldmath$-$}
\Text(25,65.5)[c]{\Large\boldmath$-$}
\BCirc(-13,38)3
\Text(-13,38)[c]{\Large\boldmath$-$}
\BCirc(2,14)3
\Text(2,14)[c]{\Large\boldmath$-$}
\BCirc(2,43)3
\Text(2,43)[c]{\Large\boldmath$+$}
\BCirc(17,32)3
\Text(17,32)[c]{\Large\boldmath$+$}
\BCirc(33,50)3
\Text(33,50)[c]{\Large\boldmath$+$}
\BCirc(58,58)3
\Text(58,58)[c]{\Large\boldmath$-$}
\BCirc(48,65)3
\Text(48,65)[c]{\Large\boldmath$+$}
\BCirc(50,33)3
\Text(50,33)[c]{\Large\boldmath$-$}
\BCirc(70,35)3
\Text(70,35)[c]{\Large\boldmath$+$}
\BCirc(17,62)3
\Text(17,62)[c]{\Large\boldmath$+$}
\Line(14,62)(13,64)
\Line(13,64)(10,61)
\Line(10,61)(7,63)
\Line(7,63)(4,61)
\Line(4,61)(1,63)
\Line(1,63)(-2,61)
\BCirc(-24.5,26)3
\Text(-24.5,26)[c]{\Large\boldmath$+$}
\Line(-21.5,26)(-20.5,28)
\Line(-20.5,28)(-17.5,25)
\Line(-17.5,25)(-14.5,27)
\Line(-14.5,27)(-11.5,25)
\Line(-11.5,25)(-8.5,27)
\Line(-8.5,27)(-5.5,25)
\BCirc(33,20)3
\Text(33,20)[c]{\Large\boldmath$+$}
\Line(36,20)(37,22)
\Line(37,22)(40,19)
\Line(40,19)(43,21)
\Line(43,21)(46,19)
\Line(46,19)(49,21)
\Line(49,21)(52,19)
\BCirc(49,45)3
\Text(49,45)[c]{\Large\boldmath$+$}
\Line(51,46.5)(52,49)
\Line(52,49)(55.5,48)
\Line(55.5,48)(57,51)
\Line(57,51)(60.5,50)
\Line(60.5,50)(62,53)
\Line(62,53)(65,51.5)
\Line(-1,48)(5,48)
\Line(-1,48)(0,49)
\Line(-1,48)(0,47)
\Line(5,48)(4,49)
\Line(5,48)(4,47)
\Text(2,51)[c]{\Large\boldmath$a_c$}
\Line(80,3)(80,77)
\Line(80,3)(79,4)
\Line(80,3)(81,4)
\Line(80,77)(79,76)
\Line(80,77)(81,76)
\Text(85,35)[c]{\Large\boldmath$L$}
\Line(18,45)(18,50)
\Line(18,45)(17,46)
\Line(18,45)(19,46)
\Line(18,50)(17,49)
\Line(18,50)(19,49)
\Text(15,47.5 )[c]{\Large\boldmath$b$}
\Line(20,80)(30,80)
\Line(20,80)(21,81)
\Line(20,80)(21,79)
\Line(30,80)(29,81)
\Line(30,80)(29,79)
\Text(25,85)[c]{\Large\boldmath$a_p$}
\Line(25,1)(33,1)
\Line(25,1)(26,2)
\Line(25,1)(26,0)
\Line(33,1)(32,2)
\Line(33,1)(32,0)
\Text(29,-1)[c]{\Large\boldmath$a$}
\end{picture}
\end{center}
\vspace*{0.5cm}
\begin{minipage}{0.48\textwidth}
\caption{A polyion of (cylindric) diameter $a_p$ 
and length $L \gg a$ surrounded by spherical counterions and 
coions of diameter $a_c$, and flexible surfactant molecules. The 
charge spacing on the polyion is $b \equiv L/Z$, and the radius of 
the exclusion cylinder is $a \equiv (a_p + a_c)/2$.}
\label{Fig.1}
\end{minipage}
\end{figure}

The strong electrostatic interaction between the polyions, the counterions, and 
the surfactants leads to formation of clusters each made of {\bf one} polyion, $n_B$ 
counterions, and $m_B$ surfactants. In what follows we shall neglect the effects of polydispersity in cluster sizes, since it can be shown not to significantly affect the final results \cite{LevBarb,Rod}. The counterion and surfactant association 
with the polyions reduces the number of free entities. Charge conservation implies

\begin{eqnarray}
\label{e1}
& & \rho_+ = \rho_{count} - n_B \rho_p \, , \\
& & \rho_- = \rho_{coion} \, , \\
& & \rho_s^+ = \rho_{s} - m_B \rho_p \, ,
\end{eqnarray}
 
\noindent
where $\rho_+$ is the density of free counterions, $\rho_-$ is 
the density of coions, and $\rho_s^+$ is the density of free amphiphiles.

\section*{\bf The Theory}

The main task of the theory is to determine the characteristic 
cluster size, i.e., to find the values of $n_B$ and $m_B$. In order to 
achieve this goal the appropriate Helmholtz free energy has to be 
constructed and minimized. The free energy can be decomposed into an 
electrostatic and an entropic contributions. The electrostatic contribution 
arises due to the {\it polyion-counterion-surfactant}, the {\it polyion-polyion}, and 
the {\it counterion-coion-surfactant} interactions. The entropic part is the result 
of mixing of various species \cite{LevBarb,Rod}.

\begin{figure}[h]

\begin{center}
\epsfxsize=8.cm
\leavevmode\epsfbox{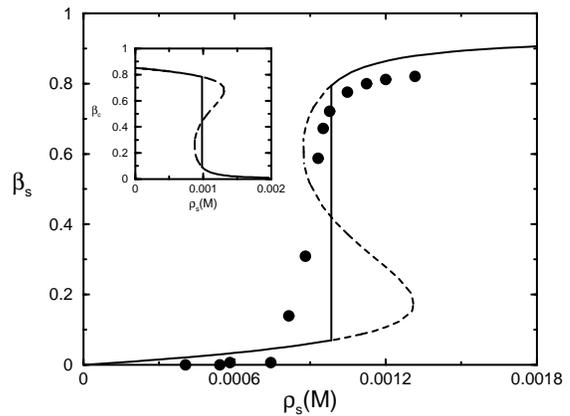}
\end{center}
\vspace*{0.5cm}
\begin{minipage}{0.48\textwidth}
\caption{The DNA-DoTAB binding isotherms, $ \chi=-4 k_BT$, $Z=440$, and $\xi=4.17$ for DNA at room temperature. The diameters of the polyions and the counterions are $27$ \AA \, and $7.04$ \AA, respectively. The size of surfactant molecule is $n_s=13$. The concentrations of DNA and of added salt are $2 \times 10^{-6}$M and $18$mM, respectively. Note that at the transition the condensed counterions are replaced by the ionic surfactants. The transition is found to be of the first order with the vertical line locating the point at which two local minima of the free energy become equal. The solid circles are the experimental data from Ref.[$13$].}
\label{Fig.2}
\end{minipage}
\end{figure}
\noindent

The {\it polyion-counterion-surfactant} and the {\it counterion-coion-surfactant} 
contributions can be obtained in the spirit of DH theory \cite{DH,LevBarb,Rod}. As 
a counterion or a surfactant associates with a polyion it neutralizes one of its 
charged groups. Hence, the effective charge per unit length of a cluster, made of 
$n_B$ bound counterions and $m_B$ bound surfactants, is $\sigma_{cl}=-q(Z-n_B-m_B)/L$. Let us fix one such cluster and ask what is the potential that it feels due to the electrostatic interactions with the other entities. In order to answer this question it is necessary to solve the Poisson equation, $\nabla^2 \Phi^{(cl)} = - 4 \pi \rho_q / D$. Due to the hard core exclusion, for $r<a \equiv (a_p + a_c)/2$ the charge distribution can be approximated as,

\begin{equation}
\label{e2}
\rho_q = \frac {\sigma_{cl}} {2 \pi} \frac {\delta(r)} {r} \, .
\end{equation}

\noindent

For $r>a$, in the spirit of DH theory, we shall assume that

\begin{eqnarray}
\label{e3}
\rho_q & = & -(Z-n_B-m_B) q \rho_p + q \rho_+ e^{-\beta q \Phi^{(cl)}(r)} + \nonumber \\
& & - q \rho_- e^{ +\beta q \Phi^{(cl)}(r) } + q \rho_s^+ e^{-\beta q \Phi^{(cl)}(r)} \, 
, \,
\end{eqnarray} 

\noindent


%
where $\beta = 1 / (k_B T)$. Upon linearization, the Poisson-Boltzmann 
equation can be easily solved to yield \cite{LevBarb,Rod}

\begin{eqnarray}
\label{e4}
\Phi_{in}^{(cl)} & = & - \frac {2 \sigma_{cl}} {D} \, \ln (r/a) + \frac {2 \sigma_{cl}} {D} \, \frac {K_0(\kappa a)} {\kappa a K_1(\kappa a)}, \, r<a \, , \\
\Phi_{out}^{(cl)} & = & \frac {2 \sigma_{cl}} {D} \, \frac{K_0(\kappa r)} {\kappa a K_1(\kappa a)} , \, r>a \, ,
\end{eqnarray} 

\noindent
where $(\kappa a)^2 \equiv 4 \pi \rho^*_1/T^*$, $\rho_1 \equiv \rho_+ 
+ \rho_- + \rho_s^+$, and the reduced density and temperature are 
respectively $\rho_i^*=\rho_i a^3$ and $T^*=Dak_BT/q^2$, while 
$K_n(x)$ are the n-order modified Bessel functions of second kind. It is important to recall that the linearization of the Poisson-Boltzmann 
equation is justified by the renormalization of polyion charge through 
formation of clusters \cite{LevBarb,Rod,AlCh,FisLev}.

In terms of this potential, the electrostatic energy of a cluster is

\begin{equation}
\label{e6}
U^{(cl)} = \frac {1} {2} \int \rho_q \, \Delta \Phi^{(cl)} \, d^3r \, ,
\end{equation} 

\noindent
with $\Delta \Phi^{(cl)} = \Phi^{(cl)}_{in} + (2 \sigma_{cl}/D) \ln (r/a)$, for $r<a$;
 $\Delta \Phi^{(cl)} = \Phi^{(cl)}_{out}$, for $r>a$. 
That is, we subtract the logarithmic potential produced by a
line of charge, since it will only contribute to the self energy of
a cluster. The electrostatic {\it free energy density}, $f \equiv -F/V$,
for the polyion-counterion-surfactant interaction is obtained through the
Debye charging process, where all the particles are charged from $0$ to
their final charge \cite{DH,Mar},

\begin{eqnarray}
\label{e7}
\beta f^{pcs} & = & -  \rho_p \, \int^{1}_{0} \, \frac {2 \, \beta U^{(cl)}(\lambda q n_B, \lambda q m_B, \lambda q Z)} {\lambda} \, d \lambda \nonumber \\
& = & - \rho_p (Z-n_B-m_B)^2 \frac {(a/L)} {T^* (\kappa a)^2} \times \nonumber \\
& & \times \left\{- 2 \ln \left[ \kappa a K_1(\kappa a) \right] + I(\kappa a) - \frac {(\kappa a)^2} {2} \right\} \, ,
\end{eqnarray}

\noindent
%
where

\begin{equation}
\label{e8}
I(\kappa a) \equiv \int^{\kappa a}_{0} dx \frac {x K_0^2(x)} {K_1^2(x)} \, .
\end{equation} 

\noindent

The electrostatic correlational free energy arising from the 
interactions between the free counterions, coions, and free 
surfactants is obtained using the usual Debye-H\"uckel theory \cite{DH,FisLev},

\begin{equation}
\label{e9}
\beta f^{ccs} = \frac {1} {4 \pi a_c^3} \left[ \ln \left( 1+\kappa a_c \right) - \kappa a_c + \frac {(\kappa a_c)^2} {2} \right] \, .
\end{equation} 

\noindent

For sufficiently large separations the effective electrostatic potential 
of interaction between two clusters separated by a distance $r$ is \cite{Rod,Ons}

\begin{equation}
\label{e10}
V_{pp}(r) = \frac {2 \pi \sigma_{cl}^2}  {D \kappa \sin \theta} \frac {\exp \left( - \kappa r \right) } {(\kappa a)^2 K_1^2 (\kappa a)} \, ,
\end{equation}

\noindent
where $\theta$ is the angle between two complexes. The short-ranged 
nature of the effective cluster-cluster interaction allows us to write 
its contribution to the free energy as a second virial term, averaged over 
the relative angle sustained by two macromolecules,

\begin{eqnarray}
\label{e11}
\beta f^{pp} & = & - \frac {1} {2} \, \rho_p^2 \, \left\langle \int d^3r 
\beta V_{pp} (r) \right\rangle_{\theta} \nonumber \\
& = & - \frac {2 \pi} {T^*} \frac {\exp \left( -2 \kappa a \right) } {(\kappa a)^4 K_1^2(\kappa a) a^3} (Z-n_B-m_B)^2 {\rho_p^*}^2 \, .
\end{eqnarray}

\noindent




The entropic (mixing) free energy is obtained using the ideas derived 
from the Flory theory \cite{Flo}. In general 
$f^{ent} = \sum_i  f_i^{ent}$, where $f_i^{ent}$ is the 
entropic contribution of each specie $i$. For free counterions and coions,

\begin{equation}
\label{e12}
\beta f_{\pm}^{ent}=\rho_{\pm} - \rho_{\pm} \ln \phi_{\pm} \, ,
\end{equation}

\noindent
where $\phi_{\pm} = (\pi \rho^*_{\pm} /6) (a_c / a)^3$ are the volume 
fractions occupied by free counterions and coions. For flexible surfactant 
chains the entropic free energy is \cite{Flo}

\begin{equation}
\label{e13}
\beta f_s^{ent}= \rho_s^+ - \rho_s^+ \ln \left[ \phi_s^+ / n_s \right] \, ,
\end{equation}

\noindent
where the volume fraction of surfactant is

\begin{equation}
\label{e14}
\phi_s^+ = n_s \frac {\pi {\rho_s^+}^*} {6} \left( \frac {a_c} {a} \right)^3 \, .
\end{equation}

\noindent
Finally, for complexes made of one rigid polyion, $n_B$ counterions, and 
$m_B$ surfactants, we find

\begin{equation}
\label{e15}
\beta f_{cl}^{ent}(\rho_p) = \rho_p - \rho_p \ln \left[ \frac {\phi_{cl} (Z + n_B + m_B)} {(Z + n_s m_B + n_B) \zeta _{cl}} \right] \, ,
\end{equation}

\noindent
with

\begin{equation}
\label{e16}
\phi_{cl} = \pi \rho^*_p \left[ \frac {1} {4 (a/L) } \left( \frac {a_p} {a} \right)^2 + \frac {1} {6} (n_s m_B + n_B) \left(\frac {a_c} {a} \right)^3 \right] \, ,
\end{equation}

\noindent
and, $\zeta_{cl}$, the internal partition function of a $(n_B,m_B)$-complex,

\begin{equation}
\label{e16a}
\zeta_{cl} = Tr e^{ - \beta H [\sigma_c(t), \sigma_s(t)] } \, .
\end{equation}

\noindent
The trace is taken over all possible configurations of $n_B$ counterions 
and $m_B$ surfactants associated to a polyion. The occupation 
variables, $\sigma_c(t)$ and $\sigma_s(t)$, are such that $\sigma_c(t)=1$ if 
the monomer $t$ of the polyion is occupied by a condensed counterion, and 
$\sigma_c(t)=0$ if no counterion is associated at $t$. The occupation variable, 
$\sigma_s(t)$, behaves in the same way, but for an association with surfactants. The 
Hamiltonian can be written as
\end{multicols}

\begin{equation}
\label{e16b}
H = \frac {q^2} {2} \sum_{t_1\neq t_2} \frac {[-1+\sigma_c(t_1)+\sigma_s(t_1)] 
[-1+\sigma_c(t_2)+\sigma_s(t_2)]} {D |r(t_1)-r(t_2)|} + \frac{\chi} {2} 
\sum_{<t_1\neq t_2>} \sigma_s(t_1) \sigma_s(t_2) \, .
\end{equation}

\noindent
\vspace{-0.5\baselineskip}

\ruleright
\begin{multicols}{2}
\noindent
An implicit constraint is that each monomer can have either a counterion or 
a surfactant associated, but not both. We have also made a simplifying assumption 
that the only effect of counterion or surfactant association is a local 
renormalization of a monomer charge. Note that the first term of the Hamiltonian 
couples all the sites, since it is due to the long-ranged Coulomb potential. The 
second sum runs only over the nearest neighbors, and is related to the hydrophobic 
interaction of the hydrocarbon tails. The configurations in which agglomerates of 
surfactant molecules form are energetically favored, i.e. the hydrophobicity parameter 
is negative, $\chi < 0$.

Even this, seemingly simple one-dimensional sub-problem, is impossible to solve 
exactly due to the long-ranged nature of the Coulomb force. We will, therefore, 
resort to a mean-field bound given by the Gibbs-Bogoliubov-Feynman inequality. 
Defining $x \equiv n_B/Z$, and $y \equiv m_B/Z$ we find \cite{Rod}
\begin{eqnarray}
\label{e8a}
\ln \zeta_{cl} & \approx & - \xi \, S \, (x^2 + 2 xy + y^2 - 2x -2y) + \nonumber \\
& & - \beta \chi \, y^2 \, (Z-1) + \nonumber \\
& & - Z (1-x-y) \ln \left( 1-x-y \right) \, + \nonumber \\
& & - Z \, \left( x \ln x \, + \, y \ln y \right) \, ,
\end{eqnarray}

\noindent
%
where $S \equiv Z \, [ \psi(Z) - \psi(1)] - Z + 1$, $\psi(x)$ is the 
digamma function, and $\xi \equiv \beta q^2 / (Db)$ is the Manning parameter 
\cite{Man}. In our previous study we have numerically checked that this, indeed, is 
a good approximation \cite{Rod}.

Minimization of the total free energy, 
$f = f^{ent} + f^{pcs} + f^{ccs} + f^{pp}$, with 
respect to $n_B$ and $m_B$ allows us to determine the 
characteristic number of bound counterions, $n^*_B$, and of bound 
surfactants, $m^*_B$. We shall compare the predictions of our theory 
with the experimental measurements on DNA dodecyltrimethylamonium bromide (DoTAB)
system \cite{Daw,Shi}.  The DoTAB is a cationic surfactant
with an alkyl chain of twelve
carbons. We can estimate the  value of the hydrophobicity parameter, $\chi$,
as follows.  Consider a micell or a monolayer composed of DoTABs.  The 
hydrophobic energy required to take an alkyl chain of twelve carbons from bulk
hydrocarbon to water is measured to be approximately $20k_BT$ \cite{Isra}. 
We can interpret
this energy as derived from the favorable interaction between the adjacent surfactants.
Since each surfactant inside a micell or a monolayer has five
or six nearest neighbors, we can estimate $\chi\approx -4k_BT$.  Clearly this is
only a rough estimate but it should be sufficient to explore the ramifications
of the new theory.

\begin{figure}[h]

\begin{center}
\epsfxsize=8.cm
\leavevmode\epsfbox{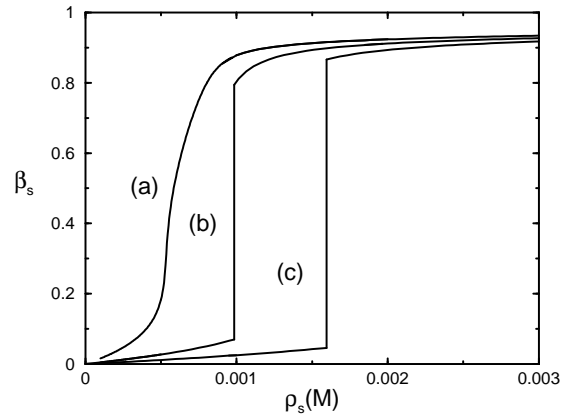}
\end{center}
\vspace*{0.5cm}
\begin{minipage}{0.48\textwidth}
\caption{The DNA-DoTAB binding isotherms for various 
concentrations of added salt: (a) 5 mM, (b) 18 mM, (c) 40 mM. The other experimental parameters are the same as in Fig.$2$. Note the 
change in the order of transition as it passes from continuous to 
discontinuous with an increase in the density of monovalent salt.}
\label{Fig.3}
\end{minipage}
\end{figure}
\noindent

In discussions of adsorption it is traditional to 
define binding fractions, $\beta_c \equiv n^*_B/Z$ and $\beta_s \equiv m^*_B/Z$. In 
Fig.$2$ we present the binding isotherms of DNA with  
dodecyltrimethylamonium bromide, and 
compare it with the experimental data of \cite{Daw}. It is evident that 
the agreement is quite good, without any fitting parameters! We note, however, that at densities of 
monovalent salt used in experiment, our theory predicts a first-order 
transition, while the experimental data is more consistent with a second-order 
transition. In Fig.$3$ we demonstrate that as the concentration of 
monovalent salt is lowered the transition becomes continuous. 

\section*{\bf Conclusion}

We have presented a mean-field theory of polyelectrolyte-ionic-surfactant solutions. Although quite simple, our theory manages to capture the essential 
physics of the problem. The most nontrivial aspect of 
polyelectrolyte-ionic-surfactant complex formation is that it occurs at 
extremely low densities, about a factor of twenty lower than the critical 
micell concentration (CMC) of pure amphiphile. This should be compared to 
the interaction of nonionic polymer with surfactant, in which case the 
binding transition happens at densities close to the CMC. Our theory 
explains this dichotomy in terms of strong electrostatic interactions which, in 
addition to hydrophobic forces, govern the polyelectrolyte-surfactant 
complex formation.

\section*{\bf ACKNOWLEDGMENTS}

P.K. would like to thank Dr. M\'ario  N. Tamashiro for the help 
with numerical methods. Y. L. is grateful to Prof. K. A. Dawson for 
bringing the importance of DNA-surfactant interaction to his 
attention, and for hospitality during his visit at the University 
College Dublin. Y. L. would also like to acknowledge helpful conversations 
with Dr. Gorelov, regarding the experimental procedure. This work was 
supported in part by CNPq - Conselho Nacional de
Desenvolvimento Cient\'{\i}fico e Tecnol\'ogico and FINEP -
Financiadora de Estudos e Projetos, Brazil.


\end{multicols}
\end{document}